\documentclass{pos}

\title{Parton distributions: HERA--Tevatron--LHC}

\ShortTitle{Parton distributions: HERA--Tevatron--LHC}

\author{\speaker{Graeme Watt}\\
        Theory Group, Physics Department, CERN, CH-1211 Geneva 23, Switzerland\\
        E-mail: \email{Graeme.Watt@cern.ch}}

\abstract{The parton distribution functions (PDFs) are a non-negotiable input to almost all theory predictions at hadron colliders.  In this talk, I introduce PDF determination by global analysis and discuss selected topics concerning recent relevant data from HERA and the Tevatron, before giving some prospects for the LHC.  The combination of H1 and ZEUS cross sections reduces uncertainties and will be an important input to future global PDF analyses.  The theoretical description of the heavy-quark contribution to structure functions at HERA has a significant influence on predictions at the LHC.  New $W$ and $Z$ data from the Tevatron Run II provide important PDF constraints, but there are currently problems describing the latest data on the lepton charge asymmetry from $W\to\ell\nu$ decays.  The Tevatron Run II jet production data prefer a smaller high-$x$ gluon than the previous Run I data, which impacts on predictions for Higgs cross sections at the Tevatron.  It is now possible to consistently calculate a combined ``PDF+$\alpha_S$'' uncertainty on hadronic cross sections, which is around 2--3\% for the $W$ and $Z$ total cross sections at the LHC, reflecting their potential as a ``standard candle'' to measure machine luminosity.  Parton luminosity functions are useful quantities for studying properties of hadronic cross sections.  Precision measurements at the LHC will provide further constraints on PDFs as data accumulates in the early running period.}

\FullConference{XXth Hadron Collider Physics Symposium \\
		 November 16 -- 20, 2009 \\
		 Evian, France}

\begin{document}

\section{Introduction}
Protons are not elementary particles: they are made of \emph{partons} (quarks and gluons).  Parton distribution functions (PDFs) are therefore essential to relate theory to experiment at HERA, the Tevatron and the LHC~\cite{slides}.  Each PDF $f_{a/A}(x,Q^2)$ intuitively gives the \emph{number density} of partons $a$ in a hadron $A$ with momentum fraction $x$ at a hard scale $Q^2\gg \Lambda_{\rm QCD}^2$.  The ``standard'' perturbative QCD framework at hadron colliders is fixed-order collinear factorisation, which holds up to formally power-suppressed (``higher-twist'') terms $\mathcal{O}(\Lambda_{\rm QCD}^2/Q^2)$.  A hadronic cross section $\sigma_{AB}$ can be written as a sum of partonic cross sections $\hat{\sigma}_{ab}$, each expanded as a perturbative series in the running strong coupling $\alpha_S(Q^2)$, convoluted with a PDF for each hadron, i.e.
\begin{equation}
  \sigma_{AB} = \sum_{a,b=q,g}\left[\hat{\sigma}_{ab}^{\rm LO}+\alpha_S\,\hat{\sigma}_{ab}^{\rm NLO}+\alpha_S^2\,\hat{\sigma}_{ab}^{\rm NNLO}+\ldots\right]\otimes f_{a/A}(x_a,Q^2) \otimes f_{b/B}(x_b,Q^2),
\end{equation}
where $\otimes$ indicates a convolution over the momentum fraction $x_{a,b}$.  The scale dependence of the PDFs is given by the Dokshitzer--Gribov--Lipatov--Altarelli--Parisi (DGLAP) evolution equations:
\begin{equation}
  \frac{\partial f_{a/A}}{\partial \ln Q^2} = \frac{\alpha_S}{2\pi}\sum_{a^\prime=q,g} \left[P_{aa^\prime}^{\rm LO}+\alpha_S\,P_{aa^\prime}^{\rm NLO}+\alpha_S^2\,P_{aa^\prime}^{\rm NNLO}+\ldots \right] \otimes f_{a^\prime/A},
\end{equation}
while the running of $\alpha_S(Q^2)$ satisfies the renormalisation group equation.  However, the input values $f_{a/A}(x,Q_0^2)$ and $\alpha_S(Q_0^2)$ to the evolution equations are incalculable by perturbative QCD and so need to be extracted from data.  Structure functions in deep-inelastic scattering (DIS) can similarly be written in terms of perturbative coefficient functions, $C_{i,a}$, convoluted with PDFs, i.e.
\begin{equation}
  F_i\left(x,Q^2\right) = \sum_{a=q,g}\left[C_{i,a}^{\rm LO}+\alpha_S\,C_{i,a}^{\rm NLO}+\alpha_S^2\,C_{i,a}^{\rm NNLO}+\ldots\right]\otimes f_{a/A}.
\end{equation}
Since the PDFs are \emph{universal}, they can be determined from a wide range of existing data, for example, from the HERA $ep$ collider (H1 and ZEUS experiments), from fixed-target experiments in $\ell p$ and $\ell d$ scattering (BCDMS, NMC, E665, SLAC), $\nu N$ scattering (CCFR, NuTeV, CHORUS), $pp$ and $pd$ scattering (E866/NuSea), together with $p\bar{p}$ collider data from the Tevatron (CDF, D{\O}).

The paradigm for PDF determination by ``global analysis'' is to parameterise the $x$ dependence of $f_{a/A}(x,Q_0^2)$ for each flavour $a=q,g$ at the input scale $Q_0^2\sim 1$~GeV$^2$ in some flexible form, subject to number- and momentum-sum rule constraints.  The PDFs are then evolved to higher scales $Q^2 > Q_0^2$ using the DGLAP evolution equations.  The evolved PDFs are convoluted with $C_{i,a}$ or $\hat{\sigma}_{ab}$ to calculate theory predictions corresponding to a wide variety of data.  The input parameters are then varied to minimise a global goodness-of-fit measure ($\chi^2$).

The determination of parton distributions by global analysis has been an ``industry'' for more than 20 years, with regular updates as new data and new theory become available.  The first NLO fit was done by the \textbf{M}artin--\textbf{R}oberts--\textbf{S}tirling group (1987), later joined by \textbf{T}horne (1998), until the retirement of \textbf{R}oberts (2005) and the addition of G.\textbf{W}.~(2006).  The previous ``MRST'' fits have recently been superseded by the ``MSTW 2008'' (LO, NLO and NNLO) fits~\cite{Martin:2009iq}.  The other major group is ``CTEQ'' (\textbf{C}oordinated \textbf{T}heoretical--\textbf{E}xperimental Project on \textbf{Q}CD), and the latest public fits are CTEQ6L1 at LO~\cite{Pumplin:2002vw}, CTEQ6.6 at NLO~\cite{Nadolsky:2008zw}, while a NNLO fit is still forthcoming.  Other groups generally fit a more restricted range of data with fewer free parameters~\cite{JimenezDelgado:2008hf,Alekhin:2009ni,H1:2009wt}.  The NNPDF Collaboration~\cite{Ball:2008by,Ball:2009mk} use an interesting alternative approach to determine PDFs from a neural network parameterisation to avoid bias due to a particular functional form of the input.

\section{HERA}
Existing HERA data already provide one of the most important inputs to global PDF analyses, especially in providing a strong constraint on the small-$x$ gluon and sea-quark distributions.  The separate H1 and ZEUS inclusive cross-section measurements in DIS have recently been \emph{combined} to improve accuracy~\cite{H1:2009wt}.  The averaging procedure gives a large reduction in the correlated systematic uncertainties due to the different properties of the two detectors, leading effectively to ``cross-calibration'' of the systematic uncertainties between the two experiments.  These new HERA combined data will prove invaluable in the next generation of global fits.  A PDF fit \emph{only} to these HERA data has already been performed~\cite{H1:2009wt}, but with only 10 free PDF parameters used to determine the central fit compared to, for example, 28 free PDF parameters for the ``MSTW 2008'' fit, reflecting the incomplete flavour separation provided by fitting \emph{only} to the HERA inclusive data.

Heavy quarks, particularly charm, can contribute a sizeable amount to the total DIS structure function $F_2$.  There are two well-defined regions in which to calculate the heavy-quark contribution.  In the \emph{fixed flavour number scheme} (FFNS), valid for $Q^2\lesssim m_H^2$, the heavy-quark mass ($m_H$) is retained in the calculation of the hard-scattering coefficient function, and there is no heavy-quark PDF which would resum $\alpha_S\ln(Q^2/m_H^2)$ terms in a similar way as for light quarks.  In the \emph{zero-mass variable flavour number scheme} (ZM-VFNS), valid for $Q^2\gg m_H^2$, this heavy-quark PDF is introduced, and the mass dependence is neglected in the coefficient function.  A \emph{general-mass VFNS} (GM-VFNS) interpolates between these two well-defined limits, using a FFNS for $Q^2\le m_H^2$ and a ZM-VFNS for $Q^2\gg m_H^2$, although there are ambiguous $\mathcal{O}(m_H^2/Q^2)$ terms in the intermediate region of $Q^2>m_H^2$.  The calculation of $W$ and $Z$ cross sections at the LHC is directly influenced by the treatment of heavy quarks in DIS, since the relevant sea-quark PDFs are determined from HERA data.  The change from CTEQ6.1 (ZM-VFNS) to CTEQ6.5 (GM-VFNS) gave a 8\% increase in $\sigma_{W,Z}$ at the LHC.  The MRST group have used a GM-VFNS since 1998, but the change from MRST 2004 to MRST 2006 introduced the first precise definition of a GM-VFNS at NNLO, including in particular the (correct) discontinuities in the NNLO PDF evolution at $Q^2 = m_H^2$, leading to a 6\% increase in $\sigma_{W,Z}$ at the LHC.  Pre-2006 MRST \emph{NNLO} (but not NLO) PDF sets should therefore now be considered obsolete due to the incomplete heavy-flavour treatment.

Heavy-flavour structure function data from HERA are reaching impressive precision, particularly for the charm structure function $F_2^{c\bar{c}}$, where the separate H1 and ZEUS measurements have also been combined using the same procedure as for the inclusive cross sections.  For both charm and beauty structure functions, there is good agreement between the data and theoretical predictions using different varieties of GM-VFNS, with the $F_2^{c\bar{c}}$ data having some discriminating power at the lowest $Q^2$ values (but still $Q^2>m_c^2$), where the GM-VFNS predictions exhibit the largest variation.  The systematic uncertainty in the particular choice of GM-VFNS, and its effect on hadronic cross sections, remains to be fully quantified, but work is in progress towards achieving this goal.

The longitudinal proton structure function, $F_L$, was measured at HERA using data taken in the last few months of running in 2007 when the proton beam energy was lowered.  The NLO and NNLO calculations tend to undershoot the HERA data at the lowest $Q^2$ and $x$ values where the theory predictions are perturbatively unstable, while small-$x$ resummation~\cite{White:2006yh} aids the description.  Small-$x$ resummation will be important at the LHC, for example, in low-mass Drell--Yan production, where the fixed-order theory predictions are also seen to be perturbatively unstable.

\section{Tevatron}
Data from the Tevatron Run II are playing an increasing r\^ole in global PDF fits.  The primary types of data used are the $Z$ rapidity distributions~\cite{Abazov:2007jy,Aaltonen:2009pc}, included for the first time in the ``MSTW 2008'' fit~\cite{Martin:2009iq}, the $W\to\ell\nu$ charge asymmetry, and cross sections for inclusive jet production.

\subsection{$W\to\ell\nu$ charge asymmetry}
The $W$ charge asymmetry at the Tevatron, as a function of the $W$ rapidity ($y_W$), is
\begin{equation}
  A_{W}(y_W) = \frac{{\rm d}\sigma(W^+)/{\rm d} y_W\;-\;{\rm d}\sigma(W^-)/{\rm d} y_W}{{\rm d}\sigma(W^+)/{\rm d} y_W\;+\;{\rm d}\sigma(W^-)/{\rm d} y_W}\approx \frac{u(x_1)d(x_2)\;-\;d(x_1)u(x_2)}{u(x_1)d(x_2)\;+\;d(x_1)u(x_2)},
\end{equation}
where $x_{1,2}=(M_W/\sqrt{s})\,\exp(\pm y_W)$, constraining mainly the $d/u$ ratio of the proton PDFs.  However, experimentally, the $W$ rapidity cannot be directly reconstructed since the longitudinal momentum of the decay neutrino is, in general, unknown.  Therefore, the quantity which is traditionally measured instead is the \emph{lepton} charge asymmetry, as a function of the lepton pseudorapidity $\eta_\ell$, i.e.
\begin{equation}
  A_{\ell}(\eta_\ell) = \frac{{\rm d}\sigma(\ell^+)/{\rm d}\eta_{\ell}\;-\;{\rm d}\sigma(\ell^-)/{\rm d}\eta_{\ell}}{{\rm d}\sigma(\ell^+)/{\rm d}\eta_{\ell}\;+\;{\rm d}\sigma(\ell^-)/{\rm d}\eta_{\ell}}.
\end{equation}
Global PDF fits have previously used Tevatron Run I data on $A_\ell$~\cite{Abe:1998rv}.  The MSTW 2008 fit~\cite{Martin:2009iq} was the first to instead use Run II data~\cite{Acosta:2005ud,Abazov:2007pm}, provided in two $E_T^e$ bins for the case of CDF data on $A_e$~\cite{Acosta:2005ud}.  The latest D{\O} data on $A_e$~\cite{Abazov:2008qv} and $A_\mu$~\cite{D0muonNote} are badly described by current NLO PDFs, especially for $p_T^\ell>35$~GeV, while refitting the PDFs causes tension with a number of other data sets, although this tension is reduced with modified deuteron corrections.  It is not possible to describe both the D{\O} $A_e$~\cite{Abazov:2008qv} and $A_{\mu}$~\cite{D0muonNote} data simultaneously.  The effect of NNLO corrections~\cite{Melnikov:2006kv,Catani:2009sm} (or $p_T^W$-resummation, as implemented in \textsc{resbos}) is small, but acts in the right direction.  CDF have recently determined $A_W$~\cite{Aaltonen:2009ta} using a new technique to obtain the neutrino's longitudinal momentum by constraining the $\ell\nu$ mass to $M_W$.  The MSTW 2008 PDFs using \textsc{vrap}~\cite{Anastasiou:2003ds} give a good description (better than the previous MRST 2006 PDFs) of the CDF $A_W$ data, while modified fits to the new D{\O} $A_\ell$ data~\cite{Abazov:2008qv,D0muonNote} tend to undershoot the CDF $A_W$ data.  Before the new precise $A_\ell$ data can be usefully included in global PDF fits, more work is needed to qualify and resolve the apparent discrepancies between (i)~CDF and D{\O} data, (ii)~$A_e$ and $A_\mu$ data, and (iii)~data and theory.

\subsection{Inclusive jet production}
The Tevatron Run I data on inclusive jet production were included in global PDF fits up to MRST 2006 (and the current CTEQ6.6) as an important constraint on the high-$x$ gluon distribution.  The MSTW 2008 analysis~\cite{Martin:2009iq} was the first to include Run II jet data~\cite{Abulencia:2007ez,Abazov:2008hua}, finding a preference for a \emph{smaller} gluon distribution at high $x$ than that obtained with the previous Run I data.  Fitting only to Run I jet data gives a bad description of Run II jet data, and vice versa, while fitting neither gives a similar description as only fitting Run II jet data.  Some similar findings have been made by the CTEQ group~\cite{Pumplin:2009nk}, although with a little less discrepancy and change in gluon.  There is therefore some apparent inconsistency between the Run I and Run II jet data, while the Run II jet data are slightly more consistent with the rest of the data included in the global fit.  The final MSTW 2008 analysis therefore dropped the Run I jet data from the fit.  There is only a slight change in the gluon if the CDF Run II data obtained using the $k_T$ jet algorithm~\cite{Abulencia:2007ez} are replaced by the CDF Run II data obtained using the cone-based Midpoint jet algorithm~\cite{Aaltonen:2008eq}.  The new smaller high-$x$ gluon is also preferred by the D{\O} Run II dijet mass spectrum~\cite{D0dijetNote}, especially at high rapidities, where the data prefer MSTW 2008 over CTEQ6.6.

The NNLO trend is similar to NLO, with the caveat that the exact NNLO jet cross sections are unavailable, so 2-loop threshold corrections are used instead.  The smaller high-$x$ gluon (and smaller $\alpha_S$) in MSTW 2008 compared to MRST 2006 means that the predicted Higgs cross sections at the Tevatron are also smaller.  The MSTW 2008 NNLO PDFs were used for the Tevatron exclusion results from last March~\cite{HiggsMarch2009} and last November~\cite{HiggsNovember2009}, whereas previous results used MRST 2002 NNLO PDFs, which fit Tevatron Run I jet data and also had an incomplete heavy-flavour treatment.

\section{LHC}
It is common to determine the strong coupling $\alpha_S$ at the same time as the PDFs.  For example, the MSTW 2008 NNLO analysis obtained $\alpha_S(M_Z^2)=0.1171\pm0.0014$ from only experimental uncertainties~\cite{Martin:2009bu}, with an additional theory uncertainty ($\lesssim 0.003$), cf.~the Particle Data Group world average value of $\alpha_S(M_Z^2) = 0.1176\pm0.002$.  The same value of $\alpha_S$ should be used in subsequent cross-section calculations.  However, since the PDFs and $\alpha_S$ are correlated, the uncertainty on a hadronic cross section due to both PDFs and $\alpha_S$ cannot simply be obtained by adding the two separate uncertainties in quadrature.  A prescription has recently been developed~\cite{Martin:2009bu} to allow consistent calculation of the combined ``PDF+$\alpha_S$'' uncertainty on a hadronic cross section.  The additional uncertainty due to $\alpha_S$ is particularly important for processes where multiple powers of $\alpha_S$ appear at lowest-order, such as Higgs production via gluon--gluon fusion or inclusive jet production, both of which enter at $\mathcal{O}(\alpha_S^2)$ at the LHC.

The $W$ and $Z$ total cross sections at the LHC are a potential ``standard candle'' for determination of the machine luminosity.  The NNLO total cross sections using MSTW 2008 NNLO PDFs have a ``PDF+$\alpha_S$'' uncertainty of around 2--3\%, while the additional uncertainty from varying the renormalisation and factorisation scales is less than 1\%.  Dependence on other theoretical uncertainties, such as heavy-quark masses and the specific choice of GM-VFNS used in the PDF fit, is currently under study.  Most uncertainties largely cancel in the $W/Z$ and $W^+/W^-$ ratios.

The parton luminosity function, $\partial{\cal L}_{ab}/\partial M_X^2$, can be interpreted as the appropriate convolution of PDFs for production of a final state with invariant mass $M_X$ from initial-state partons $a$ and $b$.  It proves very useful when studying properties of hadronic cross sections, for example, the PDF uncertainty or the dependence on different LHC beam energies~\cite{MSTWplots,Quigg:2009gg}.

Of course, as data begins accumulating at the LHC, precision measurements will provide further constraints on PDFs.  In particular, measurement of low-mass Drell--Yan production at high rapidity by LHCb~\cite{Thorne:2008am,McNulty:2008kc} may extend the small-$x$ reach of HERA, although as already noted, useful inclusion in PDF fits may require small-$x$ resummation.

The importance of PDFs can only increase now that we have firmly entered the LHC era.

\acknowledgments
I thank Alan Martin, James Stirling and Robert Thorne for collaboration.

\end{document}